\documentstyle[11pt,psfig,fullpage]{article}

\newcommand{\discCoverFig}[2]{%
  \psfig{figure=#1,width=#2,bbllx=100bp,bblly=175bp,bburx=425bp,bbury=610bp,angle=270,clip=}
  }

\begin{document}

\title{%
  \Large Data Collection for the Sloan Digital Sky Survey
  --- A Network-Flow Heuristic
  }
\author{%
  Robert Lupton\thanks{%
    Astrophysics Department,
    Princeton University, Princeton, NJ 08540.
    E-mail: {\tt rhl@astro.princeton.edu}.
    } \\
  \and
  F. Miller Maley\thanks{%
    Mathematics Department,
    Princeton University, Princeton, NJ 08540.
    E-mail: {\tt fmaley@haverford.edu}.
    } \\
  \and
  Neal Young\thanks{%
    Computer Science Department, Dartmouth College, Hanover, NH 03755.
    Parts of this research were done at:
    AT\&T Bell Laboratories, Murray Hill, NJ 07974;
    the School of ORIE, Cornell University, Ithaca NY 14853
    on \'Eva Tardos' NSF PYI grant DDM-9157199; and
    the Dept's of Astrophysics and Computer Science, Princeton University.
    Corresponding author.    E-mail: {\tt ney@cs.dartmouth.edu}.
    }
  }
\date{}

\maketitle

\pagestyle{myheadings}
\markboth{Robert Lupton, F. Miller Maley and Neal Young}{%
  Data Collection for the Sloan Digital Sky Survey
  --- A Network-Flow Heuristic}              
                                        

\begin{abstract}
  This paper describes an NP-hard combinatorial optimization problem
  arising in the Sloan Digital Sky Survey
  and a practical approximation algorithm
  that has been implemented and will be used in the Survey.
  The algorithm is based on network flow theory
  and Lagrangian relaxation.
\end{abstract}

\section{The Sloan Digital Sky Survey}
\begin{quote}
  \small
  
  ``The Sloan Digital Sky Survey
  [is] a joint project of the Astrophysical Research Consortium.
  ...
  The goal of the project, which is scheduled to begin in 1997
  and take five years, is to make a much better map of the universe
  than is currently available.
  The volume of the universe to be surveyed will be 100 times larger
  than the volume of previous surveys.
  The number of galaxies with known distances is expected to increase
  by a factor of 100 to 1,000,000 galaxies
  and the number of quasars to increase to 100,000.

  ``The Sloan Foundation ... has contributed \$8 million
  to the \$18 million capital costs of the project. ...

  ``In order to do the survey, ARC is designing and building a special purpose
  2.5 meter (100-inch) telescope at its Apache Point Observatory. ...

  ``[The Sky Survey will proceed in two phases.
  In the first phase, a two-dimensional map of the sky will be made.
  For the second phase, the] million brightest stars
  and the one hundred thousand brightest quasars will be selected
  for spectroscopic analysis from the two-dimensional map...

  \hfill \em \cite[Savani, 1994]{Savani}
\end{quote}
To gather the spectroscopic data in the second phase,
the telescope will be pointed repeatedly at the sky
to take a series of ``snapshots''.
Each snapshot will capture data for up to $660$
galaxies and quasars in the circular portion of the sky
visible through the telescope.
For each captured galaxy, light from that galaxy will enter
the telescope and travel through an optical fiber to a spectral analyzer.
The optical fibers (one for each galaxy) will be held in place
by a ``plug plate'' drilled to hold the up to $660$ fibers,
each aligned to accept the light of its respective galaxy
\cite{Crease}.

\subsection{A Capacitated Covering Problem. }
\label{problemsec}
The second phase of the survey is expected to cost
on the order of \$4-5 million.
This cost will depend primarily on the number of snapshots taken.
This paper concerns the following problem:
given the ``2-dimensional'' locations of the desired galaxies,
determine a minimum-size set of snapshots that capture them.
Formally:
\begin{quote}
  \underline{Euclidean Capacitated Covering by Disks} (ECCD)

  \em
  Given a collection of points on the unit sphere,
  a radius $r$, and a capacity $c$, find a small set of discs of radius $r$
  (located on the sphere) such that each given point
  can be assigned to a disc containing it,
  with no disc being assigned more than $c$ points.
\end{quote}
The sphere corresponds to the view-sphere centered at the telescope.
The points correspond to the images of the galaxies 
projected on the view-sphere.
Each disc represents one snapshot to be taken through the telescope;
the points assigned to that disc correspond 
to those galaxies for which data will be collected in that snapshot.
The capacity $c$ is the maximum number of galaxies for which
spectral data can be gathered in a single snapshot
(due to limitations in packing the optical fibers).

\begin{figure}[t]
  \centerline{\discCoverFig{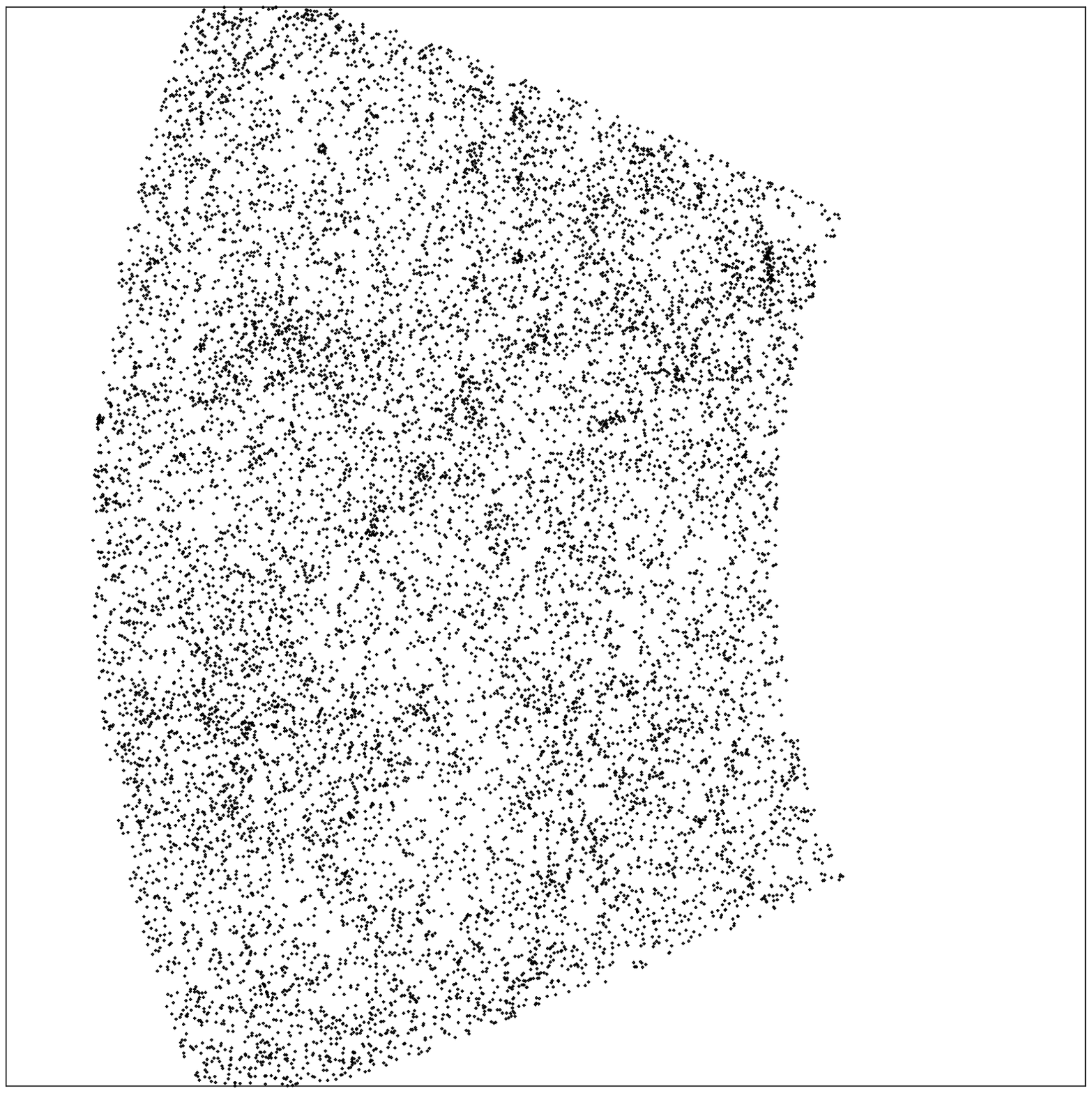}{2in}
    \discCoverFig{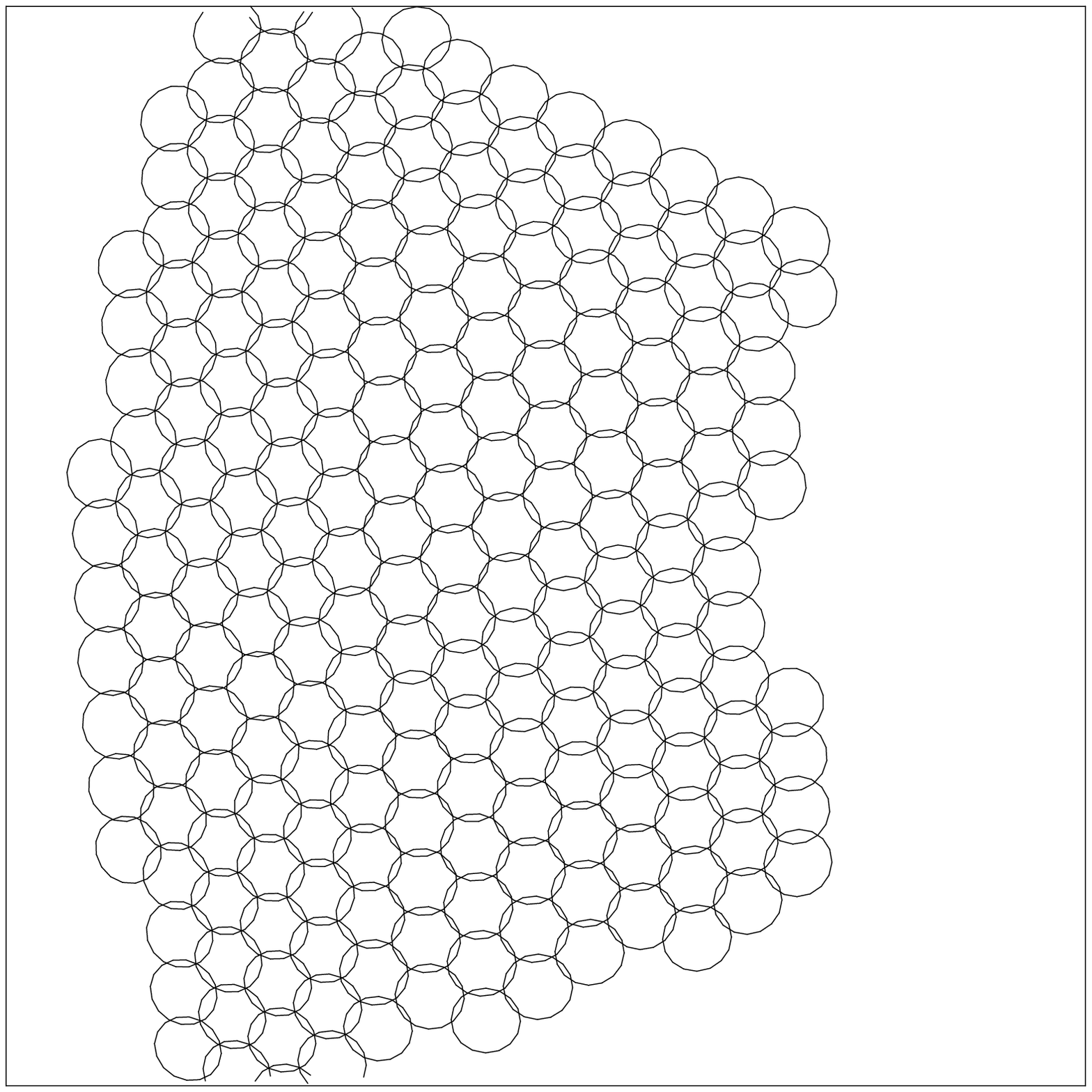}{2in}
    \discCoverFig{final.ps}{2in}}
  \caption{Sample instance (points are dark); near-uniform cover; better cover.
    This near-uniform cover is from an earlier implementation
    not using Hardin et al.'s covers, which are more uniform.
    }
  \label{fig:sample}
  \label{fig:uniform}
  \label{fig:final}
\end{figure}

The ECCD problem is NP-hard \cite{MegiddoS84}.
The instances we need to solve will have hundreds of thousands of points.
Luckily, as Figure~\ref{fig:sample} illustrates,
the instances we need to solve are nicely structured.

In this paper we describe a heuristic algorithm for the problem.
The algorithm is effective for instances arising in the Survey
and will be used for it.
The basic idea behind the algorithm is to start
with a near-uniform cover of the sphere by discs
\cite{HardinSS94} and then to iteratively improve the cover.
The key observation is that a given cover can be improved
by first solving a relaxation of the problem in which the ``point-in-disc''
constraints are replaced by penalties for assigning points
to discs not containing them,
and then moving the discs to minimize the cost of the assignment found.
The relaxed problem reduces to the minimum-cost flow problem.
In our tests, the algorithm runs in nearly linear time
and finds covers that are roughly 20\% better
than comparable near-uniform covers.

\section{Related Work}
The NP-completeness of the variant when the points lie in the plane
was proven by Megiddo and Supowit \cite{MegiddoS84}.
The proof adapts easily to our problem.
The NP-completeness of the planar problem when the discs are required
to be centered on the given points was proven by
Marchetti-Spaccamela \cite{Marchettispaccamela81}.
When the covering regions are rings, instead of discs,
Maass \cite{Maass86} showed the problem NP-complete 
even if the points all lie on a single line.

Papadimitriou \cite{Papadimitriou81}
(improving results by Fisher and Hochbaum \cite{FisherH80}),
considered the related {\em $p$-medians} problem in the plane,
which is that of covering the given points with
$p$ discs (of arbitrary radii, but centered at $p$ of the given points)
so as to minimize the {\em sum} of the disc radii.
He showed the problem to be NP-complete 
and presented average-case analyses of several algorithms.
One of the heuristics is a uniform (``honeycomb'') covering
of the points by discs,
which he shows gives a near-optimal solution with high probability
when $p$ is $\omega(\log n)$ and $o(n/\log n)$
and the points are randomly distributed in the unit square.

The problem can be modeled as a capacitated set-cover problem.
The well-known greedy algorithm of Johnson \cite{Johnson74}
and Lov\'asz \cite{Lovasz75},
as modified for the capacitated case 
by Bar-Ilan, Kortsarz, and Peleg \cite{BarilanKP93},
would yield a $\ln n$-approximate solution,
where $n$ is the number of galaxies.
This algorithm is not good enough in practice.
For this particular set-cover problem
the dual of the set system has bounded VC-dimension;
in this case an improved approximation algorithm
is known for the uncapacitated case \cite{BronnimanG95},
but, judging from a few small trials,
this algorithm does not appear to take sufficient advantage
of the structure of our problem instances to perform well in practice.

Numerous generalizations of our problem have been considered under various
names, including ``(un)capacitated facility (or plant) location,''
``$p$-centers'', and ``minimax facility location''.
These problems have been studied under various metrics
and also in general graphs.
In general, polynomial-time exact algorithms are known
only when the number of covering regions (in our case, discs) is small
(e.g., \cite{AgarwalS94})
or when the underlying metric space (or network) is tree-like
(e.g., 
\cite{MegiddoTZC81,FredericksonJ83,MegiddoT83,GurevichSV84,HeY90,ErkutFT92}).
Generally, these algorithms are for uncapacitated problems.

There is a large literature on these problems in Operations Research.
Relevant books include
\cite{LoveMW88,NemhauserW88,FrancisMW91,Francis90}.
Much of this research has concentrated 
on adapting integer-programming techniques 
to fairly general formulations of the problem.
For example, recent works on the Capacitated Facility Location Problem
(a generalization of our problem to arbitrary networks)
include \cite{CornuejolsST91,Sridharan93}.
Quoting from the conclusion of ``Approximate Solutions to Large Scale
Capacitated Facility Location Problems'' (1990) \cite{Shetty90}:
\begin{quote} \small
  The problem of locating facilities has inspired a rich body of literature
  which spans well over two decades.  Numerous algorithms have been devised
  and successfully applied to problems with as many as 200 customers
  and 100 facilities.  The computational experience on larger problems,
  however, has been virtually non-existent... In the work leading to this
  paper, the objective was to develop a heuristic algorithm that can be used
  to generate effective solutions for large scale facility locations problems.
  The computational results obtained so far seem to indicate that this
  requirement can be met for problems with as many as 1000 customers
  and 100 facilities.
\end{quote}

\section{The Algorithm}\label{sec:alg}

The instances arising in the Sky Survey exhibit particular structure.
Within any given region,
the galaxies are distributed densely throughout the region,
somewhat uniformly but with clustering tendencies
and variation in density.
The density of the galaxies means that
virtually the entire region must be covered by discs.
The variation in density means that
more discs must be concentrated within densely populated regions.
As a reference point, consider the sparsest possible covering 
of the area by discs (resembling a ``honeycomb'').  
This cover provides roughly the right {\em total}\/ capacity
and does well in sparse areas,
but in dense areas does not provide sufficient capacity.
Any good solution will have to maintain a honeycomb-like structure
in sparse areas while bunching discs more densely in dense areas.

\begin{figure}[t]
  \begin{center}
    \leavevmode
    \discCoverFig{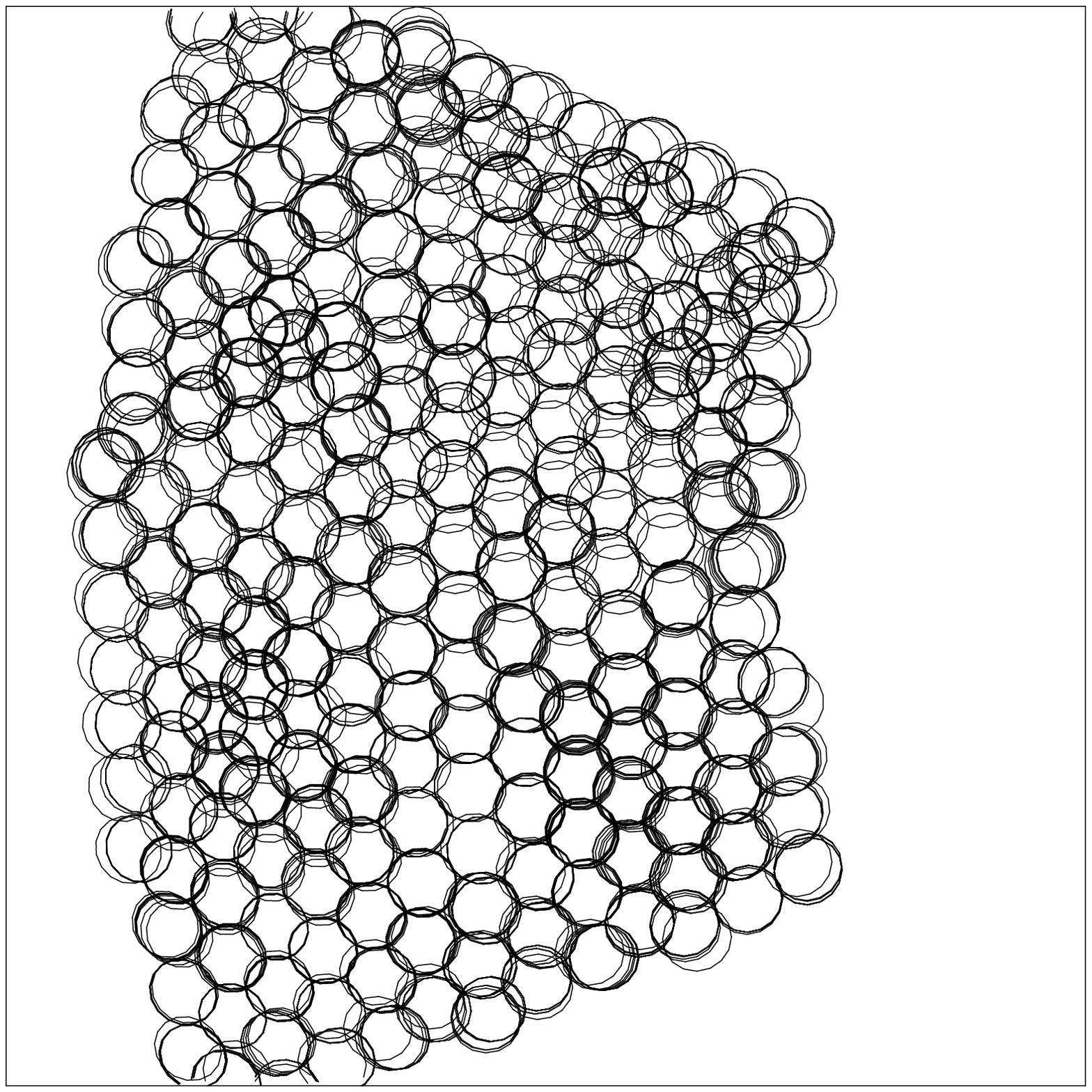}{0.5\textwidth}
  \end{center}
  \caption{Composite of intermediate covers}
  \label{fig:move}
\end{figure}

The outer loop of the algorithm does a binary search for the smallest value of
a density parameter $\delta$ that leads to success in the inner loop.  The
inner loop begins with a near-uniform cover of normalized density $1+\delta$
and iteratively improves it (see Figure~\ref{fig:uniform} for ``before'' and
``after'' covers).  Each iteration of the loop perturbs the discs, as described
below, in an attempt to improve the cover (Figure~\ref{fig:move} shows the
results of such a series of improvement steps).  If the desired coverage is
obtained, the inner loop stops (successfully).  If the perturbation ceases to
improve the cover, the inner loop stops (unsuccessfully).

Next we describe how the algorithm perturbs a given cover in order to improve
it.  We start with the observation that for a {\em given}\/ set of discs (with
known locations), the problem of finding the maximum number of galaxies that
can be assigned reduces to a generalized maximum bipartite matching problem
in a graph $G=(U,V,E)$, where the vertices in $U$ correspond to galaxies, the
vertices in $V$ correspond to the discs, and edge $(u,v)$ is present if $u$'s
galaxy is in $v$'s disc.  A maximum legal assignment of galaxies to discs then
corresponds to a maximum size set $S$ of edges such that each $u \in U$ is
incident to at most one edge in $S$ while each $v\in V$ is incident to at most
$c$ edges in $S$.

Since the latter problem reduces in a standard way
to the maximum flow problem \cite{PapadimitriouS82},
which can be efficiently solved, it follows that
for a {\em given} set of discs, one can efficiently find a
maximum legal assignment of galaxies to discs.

Of course, the maximum legal assignment may still leave many galaxies unassigned,
even though many discs are not used to capacity.
In this case, how can discs be moved to improve the coverage?
Consider the following relaxation of the problem:
\begin{quote}{
    \underline{Relaxed Problem}
    
    \em
    Given a set of discs, a set of galaxies, and a capacity $c$,
    find a {\bf minimum-penalty} assignment of the galaxies to discs
    such that no disc is assigned more than $c$ galaxies.}
\end{quote}
Here a galaxy can be assigned to a disc not containing it,
but there is a penalty for doing so that encourages assignments
of galaxies to nearby discs (details of the penalty function are
in \S~\ref{sec:impl}).

The relaxed problem can be solved efficiently (even for arbitrary penalties) by
reducing it to the assignment problem or to minimum-cost maximum flow.  We
reduced it to the latter, more general, problem in anticipation of having to
incorporate more complex constraints on the assignment (that no sufficiently
close pairs of galaxies should be assigned to the same disc) at a later point.
As described below, even the more general problem can be solved quickly enough
for our purposes.

\begin{figure}[t]
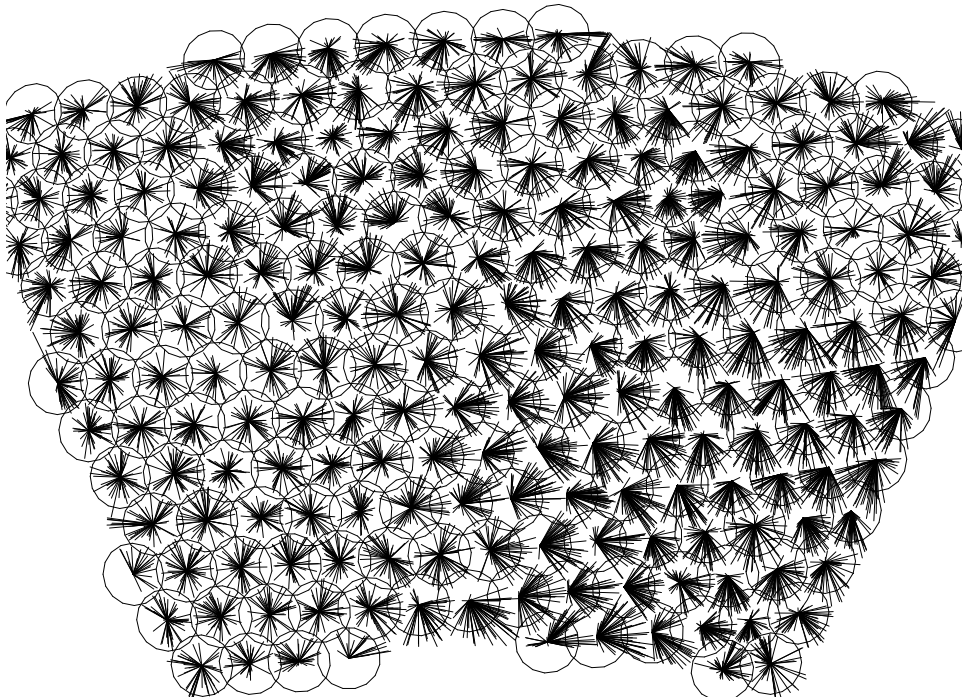

  \begin{center}
    \leavevmode
    \discCoverFig{relax.ps}{5in}
  \end{center}
  \caption{Relaxed assignment}
  \label{fig:relax}
\end{figure}

A solution to the relaxed problem will assign all galaxies to discs, but a
given disc may be assigned galaxies outside of it.  {\em The advantage of the
  relaxed problem is that a solution to it can give information about how to
  improve a given set of discs.} The intuition is that if excess demand (i.e.\ 
a high density of galaxies relative to discs) exists in one area, and excess
capacity exists in another, then a disc between the two areas will tend to be
assigned galaxies that are outside of the disc and that lie towards the area of
excess demand.  Figure~\ref{fig:relax} illustrates this.

Once a minimum-penalty solution to the relaxed problem has been found, the
algorithm moves the discs to minimize the cost of the particular assignment of
galaxies to discs specified by the minimum-penalty solution.  This problem can
be solved {\em independently} for each disc.  For a given disc, for a fixed set
of galaxies assigned to it, the sum of the penalties for those assignments is a
function $f(x,y)$ of the coordinates $(x,y)$ of the center of the disc.  As
long as the penalty function is convex and reasonably smooth, $f$ will be also.
Starting with the current location $(x_0,y_0)$ of the disc, a simple
gradient-based method (described in \S~\ref{sec:impl}) is used to find $(x,y)$
maximizing $f(x,y)$.

\newenvironment{tabAlgorithm}{
\setcounter{algorithmLine}{1}
\samepage
\begin{tabbing}
999\=\kill
}{
\end{tabbing}
}
\newcounter{algorithmLine}
\newcommand{\algline}{\\\thealgorithmLine\hfil\>\stepcounter{algorithmLine}}

\begin{figure}[tb]
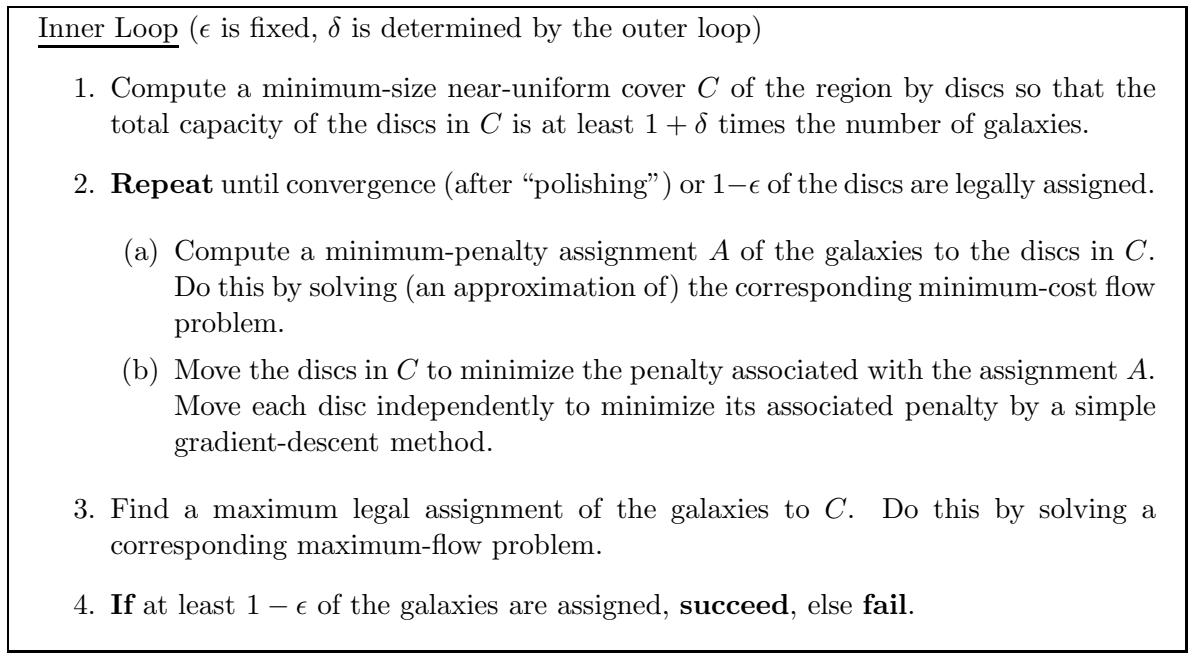

  \begin{center}
    \framebox[.95\textwidth][c]{\parbox{.9\textwidth}{
        \underline{Inner Loop} ($\epsilon$ is fixed, $\delta$ is determined by
        the outer loop)
        \begin{enumerate}
        \item Compute a minimum-size near-uniform cover $C$ of the region by
          discs so that the total capacity of the discs in $C$ is at least
          $1+\delta$ times the number of galaxies.

        \item {\bf Repeat} until convergence (after ``polishing'') or
          $1-\epsilon$ of the discs are legally assigned.

          \begin{enumerate}

          \item Compute a minimum-penalty assignment $A$ of the galaxies to the
            discs in $C$.  Do this by solving (an approximation of) the
            corresponding minimum-cost flow problem.

          \item Move the discs in $C$ to minimize the penalty associated with
            the assignment $A$.  Move each disc independently to minimize
            its associated penalty by a simple gradient-descent method.

          \end{enumerate}
        \item Find a maximum legal assignment of the galaxies to $C$.  Do
          this by solving a corresponding maximum-flow problem.

        \item {\bf If} at least $1-\epsilon$ of the galaxies are assigned,
          {\bf succeed}, else {\bf fail}.
        \end{enumerate}
        }}
  \end{center}
  \caption{
    Given a desired coverage $1-\epsilon$, where $\epsilon \ge 0$, the outer
    loop of the algorithm does a binary search for the smallest value of
    $\delta \ge 0$ such that the above inner loop succeeds.  Further details,
    including the ``polishing'' step, the ``approximation'' of the flow
    problem, and the criteria for convergence, are described in
    \S~\protect\ref{sec:impl}.}
  \label{fig:alg}
\end{figure}
This gives us the essentials of the inner loop of the algorithm.  It starts
with a near-uniform cover of some specified (normalized) density $1+\delta$.
It improves the cover by finding a minimum-penalty assignment of the galaxies
to the discs and then moving the discs to their optimal locations given that
assignment.  It continues, alternately improving the assignment and then moving
the discs, until the net penalty ceases to decrease appreciably.  At the end of
the inner loop, the algorithm finds a legal (not relaxed) assignment of
galaxies to discs maximizing the number of assigned galaxies.
Figure~\ref{fig:move} shows a sequence of covers generated by a single run of
the inner loop.

The outer loop performs a binary search to find the smallest $d$ that causes
the inner loop to successfully cover the galaxies.  The presentation here is a
slight simplification of the actual algorithm, in that the actual algorithm
uses a ``polishing'' heuristic before terminating the inner loop, and a
heuristic is applied to reduce size of the network-flow problem before solving
it.  These heuristics and other details about convergence of the inner and outer
loops, and starting conditions for the outer loop, are described in
\S~\ref{sec:impl}.

\subsection{Example Run of Inner Loop. }
The sample instance in Figure~\ref{fig:sample} contains 12642 points ---
a random $10\%$ of the points in a subregion of the sky previously scanned.
The size of this subregion is about $10\%$ of that of the region that will be
mapped by the Survey.
A uniform cover of 218 discs of capacity 60 (total capacity 13080)
allows 81\% of the galaxies to be assigned.
After 16 iterations of the inner loop of the algorithm,
the improved cover captures 97.8\% of the points.
Figure~\ref{fig:sample} shows the initial near-uniform cover
and the final cover;
Figure~\ref{fig:move} shows a composite of the successive covers.
Section~\ref{sec:perf} describes comprehensive testing of quality of solutions
given by the algorithm and its running time.

\subsection{Implementation Details. }\label{sec:impl}
For the initial near-uniform covers, we use
Hardin, Sloane, and Smith's catalogue of packings
of points on the sphere \cite{HardinSS94}.
These packings give covers of the entire sphere,
but we need a cover of only a (usually rectangular) subregion of the sky.
To prune a ``global'' cover $C$ the algorithm
first finds a maximum legal assignment of galaxies to discs in $C$,
then discards all discs having at most a few assigned galaxies.
(The cutoff for discarding a disc is chosen
so that the resulting number of discs is as desired.)

The inner loop of the algorithm is implemented in C++ using LEDA \cite{MehlhornN}
for basic data structures.
We use a scaling algorithm by Andrew Goldberg to solve
the minimum-cost flow problems \cite{Goldberg1997}.
We use TCL for the outer loop of the algorithm and to collect performance data.

\begin{figure}[tb]
  \centerline{
    \psfig{figure=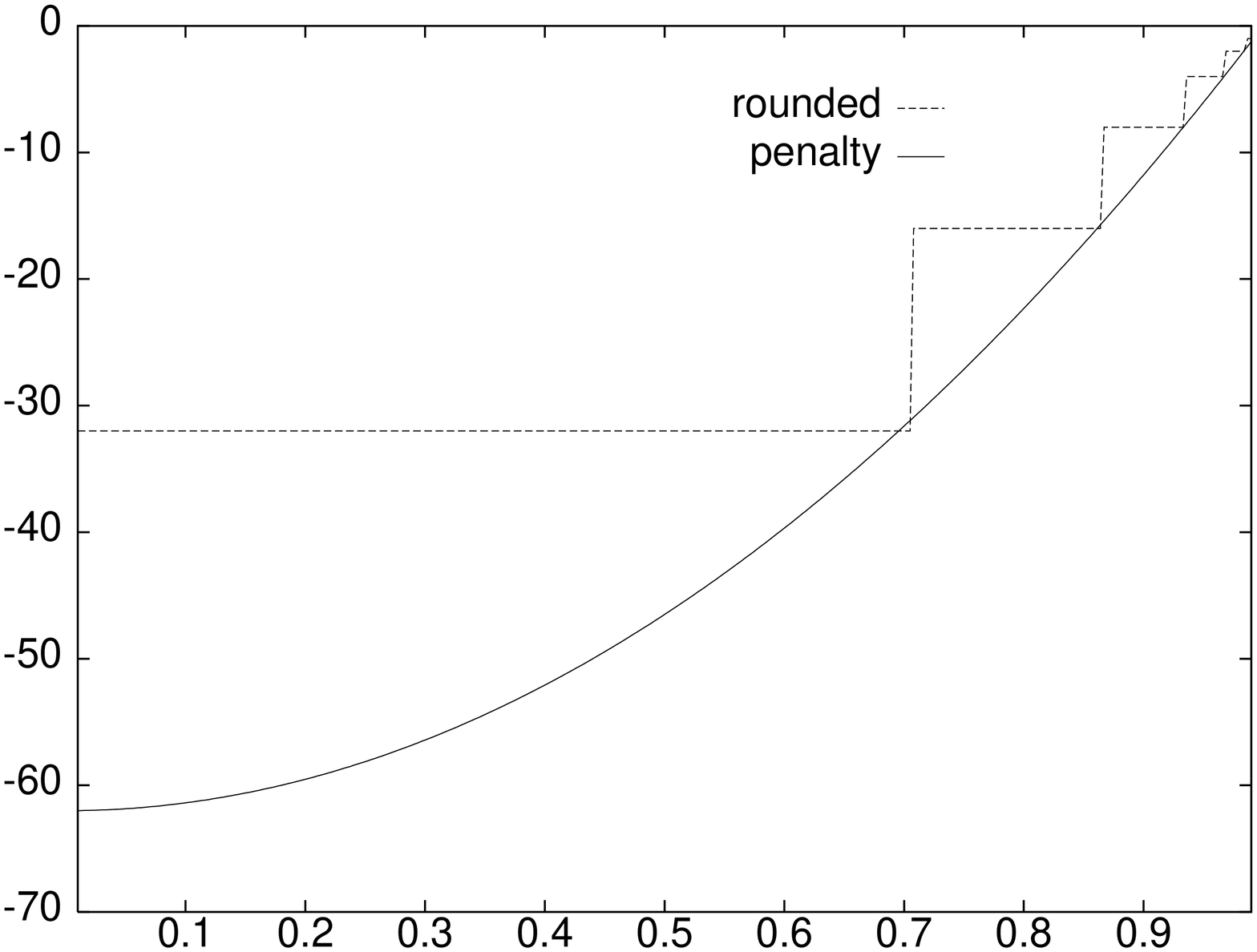,width=3.3in,angle=270}
    \psfig{figure=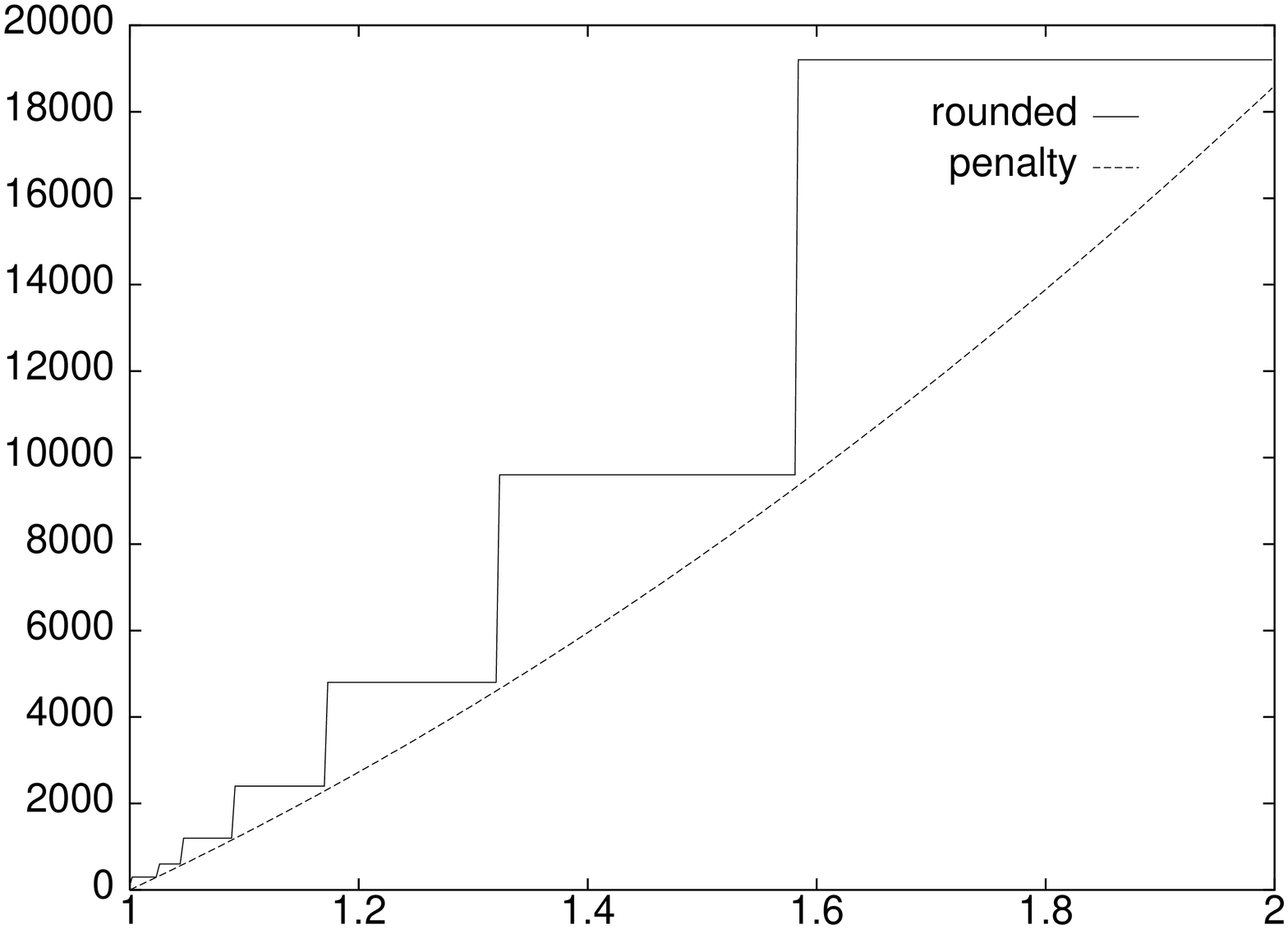,width=3.3in,angle=270}
    }
  \caption{
    The assignment penalty as a function of the distance $d$ (in disc
    radii) between the disc center and the galaxy.  The plot on the left is
    for $d \le 1$; the plot on the right is for $d \ge 1$.}
  \label{fig:cost}
\end{figure}
\paragraph{Penalty function:}
The penalty for assigning a galaxy to a disc
whose center is distance $d$ away is proportional to
$$p(d) = \cases{d^2-r^2 & if $d\le r$ \cr 100(d^2-r^2) & if $d\ge r$.}$$
Recall that $r$ is the disc radius.
When solving the relaxed problem,
the algorithm first {\em rounds} the penalties.
Rounding so that there are few distinct penalties
allows a heuristic reduction in the size of the resulting flow problem.
(This heuristic is discussed further below.)
Figure~\ref{fig:cost} shows plots of $p$ and the rounded penalties.
The rounding is chosen to preserve the distance
between the galaxy and the {\em edge} of the disc
within roughly a factor of 2.
The edge of the disc is important
because the penalty function is least smooth
for points near the edge.
The factor of 2 is somewhat arbitrary,
it was chosen to balance between
the advantages of rounding and the resulting loss of accuracy.
After rounding, only 14 or so distinct penalties (each an integer power of 2) arise.

\paragraph{Reducing the size of the flow problem:}
We expected the bottleneck in the algorithm
to be solving the minimum-cost flow problems.
To minimize this time, the algorithm uses a heuristic
to reduce the minimum-cost flow problem to a smaller,
approximately equivalent, problem.
This is the ``approximation'' of the flow problem
mentioned in the high-level description of the algorithm.
First, the algorithm only considers assigning each galaxy
to discs whose centers are within a distance of 2 disc radii,
and of these at most the three closest discs.
Second, it rounds the penalties as described above
to reduce the number of distinct penalties.
Finally, instead of having vertices for individual galaxies,
it has vertices for equivalence classes of galaxies,
where two galaxies are equivalent
if they have the same assignable discs
with the same rounded penalties.
With these heuristics,
even for very dense sets of galaxies,
the number of equivalence classes will
be proportional to the number of discs
as long as each disc intersects $O(1)$ other discs.
This is true in our case.

The precaution of using equivalence classes
turned out to be unnecessary for two reasons.
First, the average number of galaxies per equivalence class was
typically no more than $3$.
More fundamentally, solving the flow problems
was not in fact a substantial bottleneck
(see the data in \S~\ref{newsec}
and the subsequent discussion).\footnote{%
  It is conceivable that the rounding of the penalties decreased the time
  used by the minimum-cost flow algorithm, as the latter works by scaling.}

\paragraph{Constructing the flow problem:}
The algorithm stores all the discs in a two-dimensional array
so that discs near any given point can be found rapidly.
To construct the flow problem, the algorithm iterates through the galaxies.
For each galaxy, it finds the discs whose centers are within 2 disc radii.
It selects the three nearest of these discs
and computes the rounded distances to each.
These discs and their rounded distances determine
the equivalence class of the galaxy.
The equivalence class is found (or created if necessary).
From the equivalence classes, the flow network is constructed.

So that the equivalence classes can be found quickly,
each equivalence class is stored in a hash table
maintained at its nearest disc.
The hash table for a disc $D$ contains those equivalence
classes whose nearest disc is $D$.
This method preserves locality of reference.
In an earlier implementation, a single large hash table
held all the equivalence classes.
For large problems, this table was too large to fit in main memory.
This slowed the algorithm by a factor of roughly 50.

\paragraph{Moving the discs:}
After the minimum-penalty relaxed assignment is found,
recall that each disc is moved individually to minimize the penalty
associated with that disc.
The ``simple gradient-descent method'' used to do this is as follows.
To minimize $f(x,y)$, starting at a point $(x_0,y_0)$,
compute the gradient (direction of maximum rate of increase),
then move $(x,y)$ in steps of $\alpha$
(approximately $16/1000$ disc radii,
chosen to balance speed and accuracy)
in the direction opposite the gradient
until such steps ceased to decrease the value of $f(x,y)$.
Recompute the gradient at the new location
and repeat the process with $\alpha$ halved.
Continue in this fashion, halving $\alpha$ each time,
until $\alpha$ is decreased to approximately $2/1000$ disc radii.

\paragraph{Convergence and ``Polishing'':}
The outermost loop of the algorithm
does a binary search on the size of the uniform starting cover.
Within this loop, the inner loop
iteratively improves the given cover.

We describe convergence of the inner loop first.
Recall that the inner loop starts with a given cover and improves it
until the desired number of galaxies are legally covered,
or until ``convergence'' occurs.  Convergence is determined as follows:
after each iteration, if the gap between the actual number of
galaxies covered and the desired number did not decrease by at least 5\%,
then the algorithm considers the process ``stuck''.
At this point it changes the basic improvement step
(this is the ``polishing'' heuristic mentioned in the high-level descriptions
of the algorithm) as follows:
it solves the relaxed problem {\em as if} the disc radius were 2\% smaller.
It continues with this heuristic until it also becomes stuck.
Every time the process becomes stuck, the algorithm alternates
between the standard improvement step and the modified one.
If the process is ever stuck for at least two sequential rounds,
it is considered to have converged.
The purpose of the polishing heuristic
is that in the original relaxed problem,
a disc may be assigned galaxies
that are just barely outside of it at little penalty.
These galaxies cannot be legally assigned,
yet may ``hold'' discs in place in the subsequent disc-moving step.
``Shrinking'' the effective radius of the disc for a few rounds
encourages these galaxies to be assigned elsewhere.

Next we describe initial conditions and the convergence criterion for the outer loop.
The outer loop maintains a lower bound $L$ and an upper bound $U$
on the minimum sufficient cover size.
It also maintains covers $C_L$ and $C_U$ obtained by starting with a uniform
cover of size $L$ or $U$ (respectively)
and applying the basic algorithm to improve the cover until the desired
coverage is obtained or convergence occurs.
Initially $L$ and $U$ are taken to be $1.05$ and $1.15$, respectively,
times the number of galaxies divided by the capacity per disc.
The binary search maintains the invariant that $C_L$ and $C_U$ are,
respectively, insufficient and sufficient to achieve the desired coverage.
If this invariant does not hold initially, $L$ and/or $U$ are adjusted in
increments of 5\% to achieve the invariant.
The algorithm halts the binary search as soon as
the following condition ceases to be met:
$C_U$ has more than one more disc than $C_L$,
$C_U$ is at least 0.5\% bigger than $C_L$,
and $C_U$ legally covers at least 0.5\% more galaxies than $C_L$.
Once the search halts, the algorithm returns $C_U$.

\section{Performance of the Algorithm}\label{sec:perf}
We tested the running time and the quality of the solutions
found by the algorithm on sample instances.
In this section we describe the results.

The Survey will map roughly 25\% of the sky
--- the region having right ascension zero through $360$ degrees
and declination $30$ degrees through $90$ degrees.
Roughly one million galaxies will be mapped.
Because the two phases of the Survey will be pipelined
(the second will be started before the first is done),
the second phase will be done in pieces.

\begin{figure}[tb]
  \begin{tabular}{cc}
    \begin{tabular}[b]{c}
      \begin{tabular}{|c||r|r|r|} \hline
        name & r.\ ascens. & declination & galaxies \\ \hline\hline
        b & 35 to 55 & -55 to -35 & 29933 \\\hline
        c & 32 to 57 & -57 to -32 & 45344 \\\hline
        d & 30 to 59 & -55 to -30 & 52520 \\\hline
        e & 28 to 62 & -57 to -28 & 70339 \\\hline
        f & 25 to 65 & -60 to -20 & 109681 \\\hline
        g & 20 to 70 & -70 to -18 & 157126 \\ \hline
      \end{tabular}
      ~\\
      ~\\
      ~\\
      ~\\
    \end{tabular}
    &
    \psfig{figure=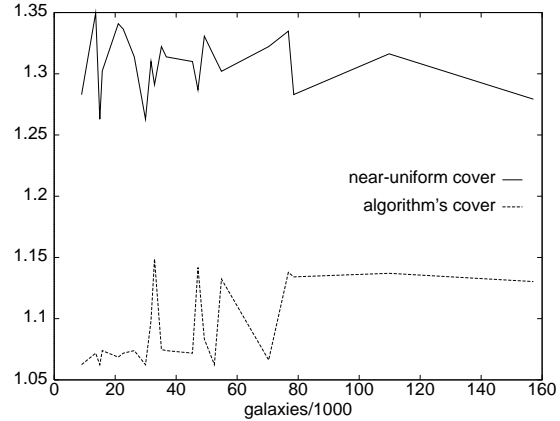,width=3in,angle=270}
  \end{tabular}
  \caption{%
    Regions from which sample instances were generated;
    number of discs needed to achieve a $98\%$ coverage
    (normalized by capacity lower bound).
    }
  \label{fig:regions}
  \label{fig:performance}
\end{figure}
We generated the problem instances from data from a region of the sky
that had been previously scanned for a different purpose.
We selected 6 subregions, and for these subregions
we generated 4 problem instances by randomly sampling
30, 50, 70 or 100\% of the galaxies.
This gave us 24 sample problems.
We took the disc radius to be 1.5 arc-seconds
and the capacity to be 600
times 0.3, 0.5, 0.7, or 1 corresponding to the sampling percentage above.
(The base capacity is 600 instead of 660 because approximately 60 points in
each disc will be reserved for quasars not in the sample.)
The largest region has an area roughly 4\% of the entire sky.
For each subregion, the right ascension and declination ranges
and the number of galaxies are shown in Figure~\ref{fig:regions}.

\subsection{Quality of solutions. }
Figure~\ref{fig:performance} illustrates the quality of the solutions
returned by the algorithm on the 24 problem instances.
The figure plots the size of the cover needed
to assign 98\% of the galaxies in each region,
normalized by dividing by the number of discs needed just to provide
enough capacity to hold 98\% of the galaxies.
The plot shows the same information for covering by near-uniform covers.
The algorithm (very roughly) requires 5\% to 15\% extra capacity,
whereas using uniform covers requires 25\% to 35\% extra capacity.

\subsection{Running time.  }
\label{newsec}
\begin{figure}[tb]
  \centerline{\psfig{figure=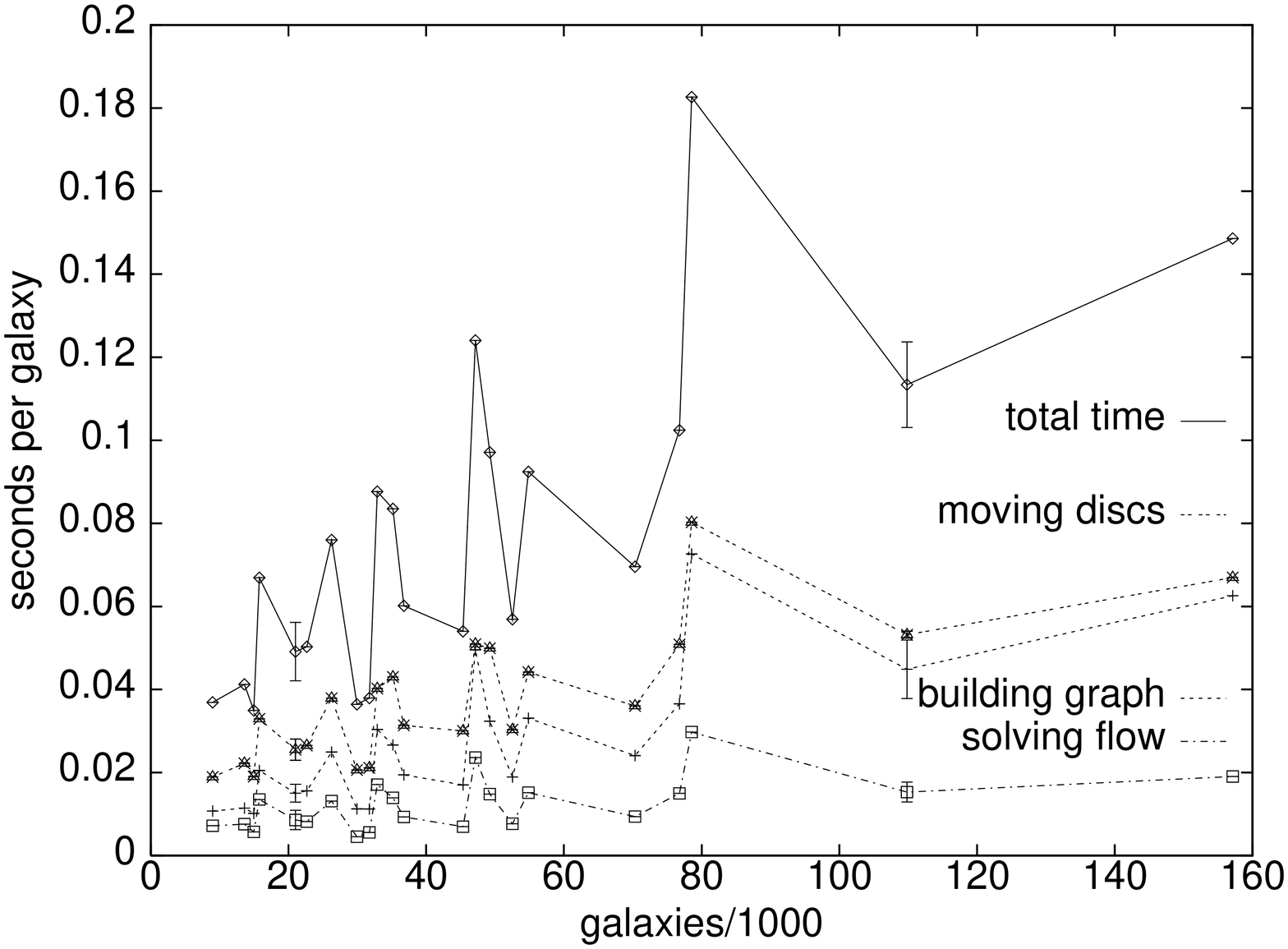,width=3.3in,angle=270}
    \psfig{figure=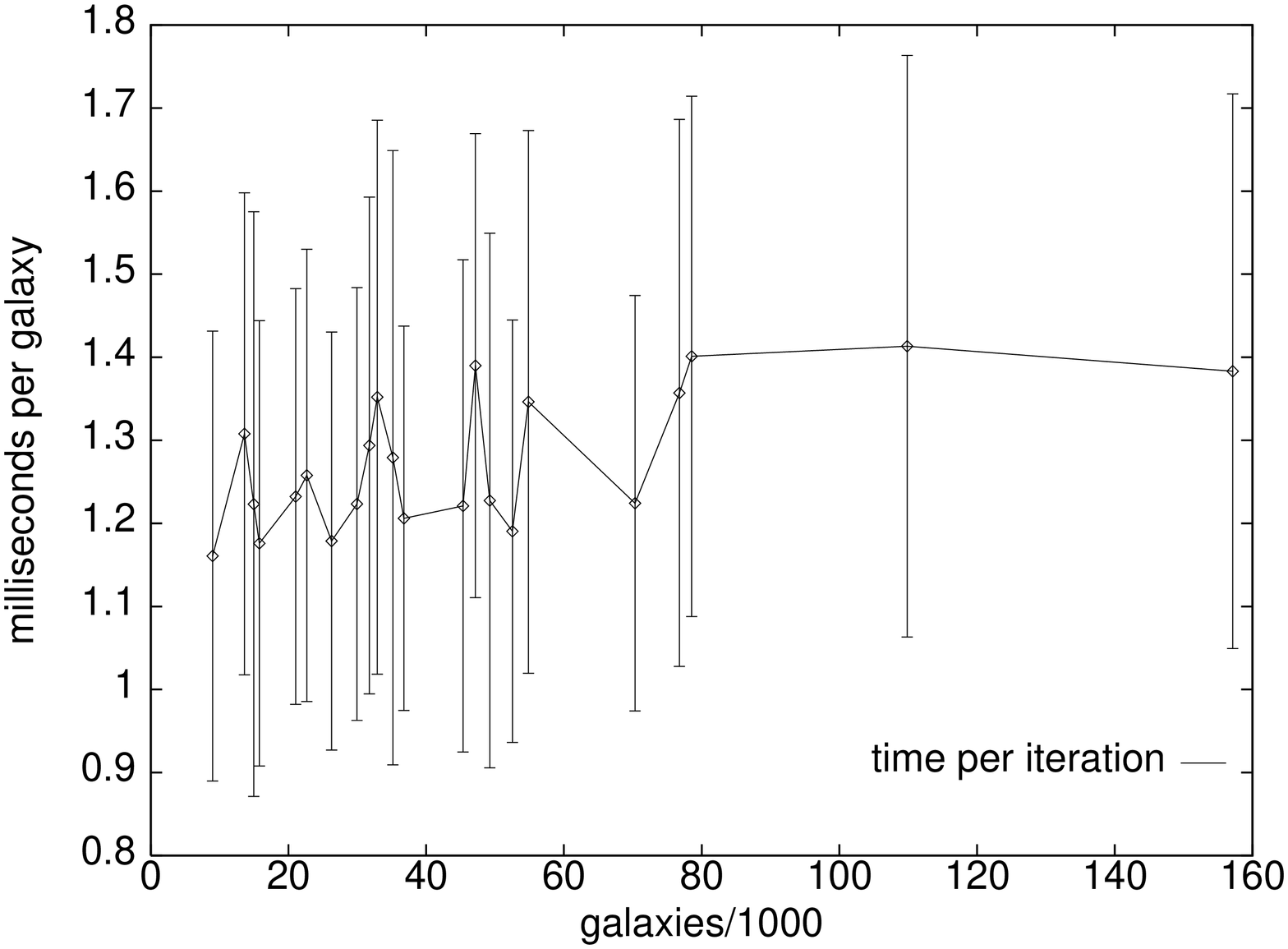,width=3.3in,angle=270}}
  \caption{Net time per galaxy and main components;
    time per galaxy per iteration.
    Each vertical bar represents a group of points with close $x$-coordinates:
    the center of the bar is the average; the endpoints are one standard
    deviation away.
    }
  \label{fig:time}
  \label{fig:time_normalized}
\end{figure}

Plots of the time per galaxy to solve each problem instance
as a function of the number of galaxies
appear in Figure~\ref{fig:time}.
This net time includes all of the iterations needed to find
the final cover for the given problem instance, including
the binary search  ``outer loop''.

The three main components of the running time are
the time building the graphs
(including finding the equivalence classes of galaxies),
the time solving the flow problems,
and the time moving the discs.
These plots show that the net running time is on the order of $0.1$
cpu seconds per galaxy ($850000$ galaxies per day),
with the three main components each taking a substantial fraction of the time.
These tests were carried out on a Silicon Graphics machine
with 6 150 MHZ processors,
a 16 Kbyte data cache,
a 1 Mbyte secondary cache, 
and 256 Mbytes of main memory.

Most of the variation in the time per galaxy is due to the number of
iterations, which varied from 30 to 100 per problem instance,
and which increases (in our implementation) with the problem size.
Figure~\ref{fig:time_normalized} plots the average time per galaxy per
iteration versus the problem size.
Each iteration represents the solution of one
relaxed problem and one perturbation of one set of discs.
The time per iteration grows linearly with the number of galaxies.
This is as expected, except for the
surprising speed of the flow computations.

We note that the binary search is fairly naive,
given that in principle a fairly precise guess about the correct size
of the starting cover could be made.
Similarly, we feel that a more careful
and less conservative estimate of convergence,
possibly interleaving the two loops in some fashion,
might be warranted.
These improvements might reduce the total number of iterations substantially.

\begin{figure}[tb]
  \centerline{\psfig{figure=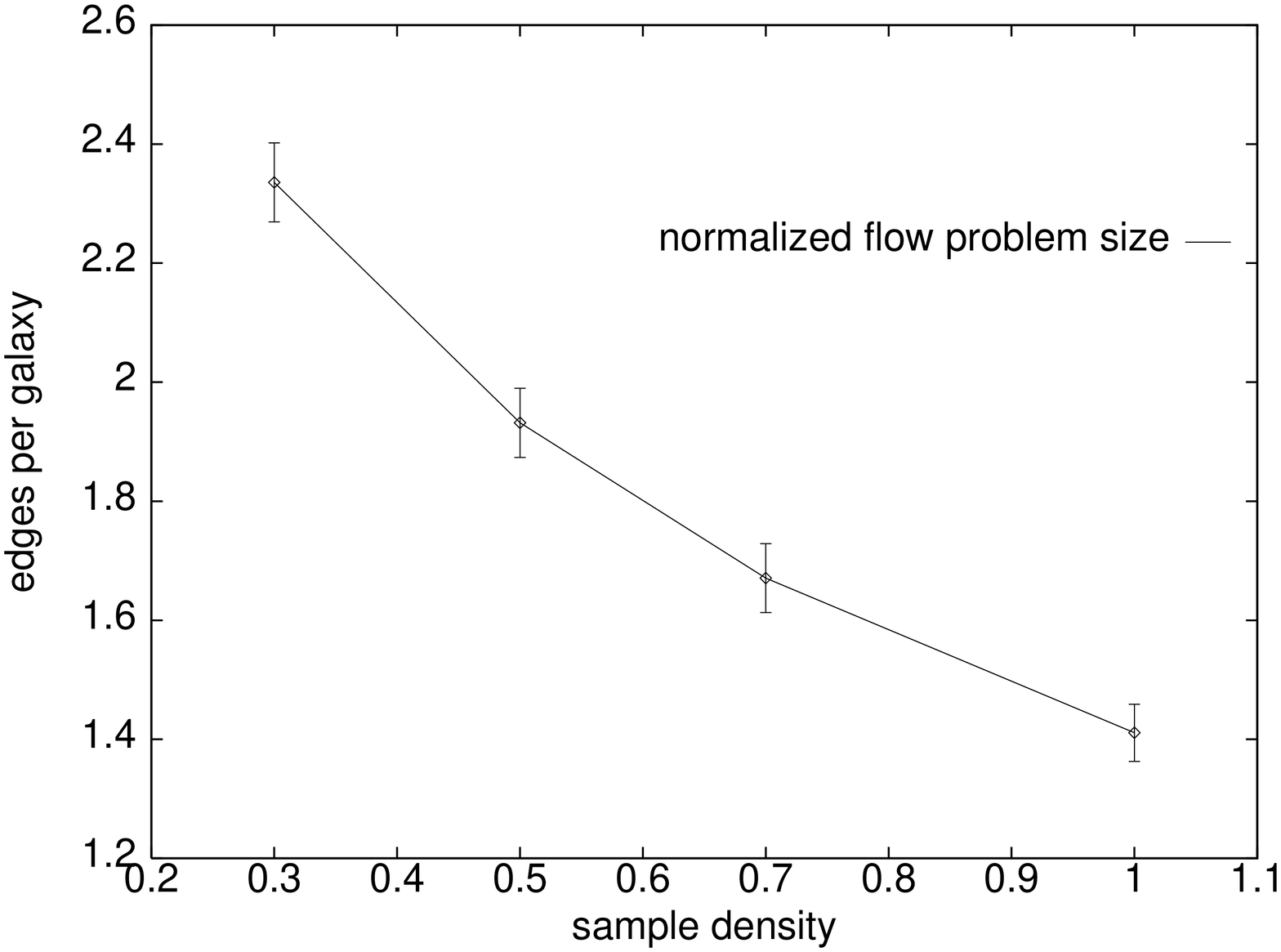,width=3.3in,angle=270}
    \psfig{figure=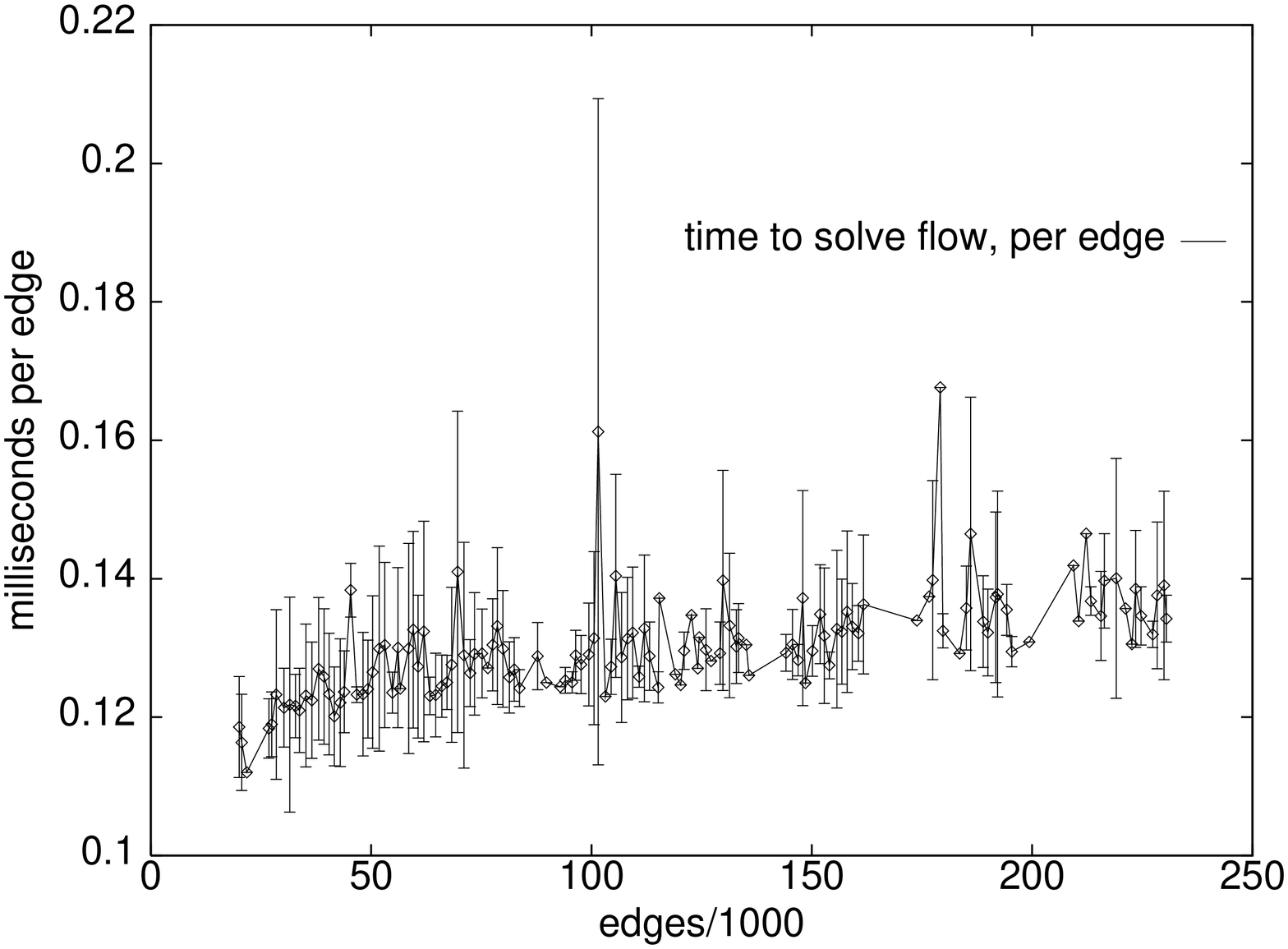,width=3.3in,angle=270}}
  \caption{Number of edges per galaxy vs.~density;
    time per edge vs.\ number of edges}
  \label{fig:flowsize}
  \label{fig:flowtime}
\end{figure}

\paragraph{Time to solve flow problems. }
The heuristics for keeping the flow problems small appear to be effective.
Figure~\ref{fig:flowsize} plots the average number of edges per galaxy
in each flow problem as a function of the sampling density of the instance.

Figure~\ref{fig:flowtime} plots the average time per edge
to solve the individual flow problems.
The time appears to grow only near-linearly with the number of edges.
Better than worst-case behavior on certain classes of problems
is not uncommon;
further the flow problems arising here are not particularly hard ones.
See \cite{DIMACS93} for computational studies related to this issue.

\section{Retrospective}
The Euclidean capacitated covering problem arising here is very natural.
Looking in the Operations Research literature, we found numerous
capacitated covering algorithms based on integer linear programming,
but these were not fast enough problems of the desired size.
The Computer Science literature had a number of efficient approximation
algorithms for covering that had provable worst-case performance guarantees,
yet these algorithms would not produce good enough solutions in practice.

Nonetheless, our final solution rests on theoretical foundations.
Our algorithm works in the spirit of Lagrangian relaxation.
We decompose the problem into two parts:
finding the cover $C$ and finding the assignment $A$.
We relax the constraints on $A$
by replacing the ``disc-containment'' constraint by a penalty function.
Then, for any given cover, finding the minimum-penalty assignment
is a tractable problem.
Likewise, for any given assignment,
finding the minimum-penalty cover is tractable.
Thus, the relaxation yields a scheme
that iteratively reduces the minimum penalty
and so drives the pair $(C,A)$ closer to feasibility.
Lagrangian relaxation is a common technique
in both the Operations Research \cite{Sridharan93}
and the Computer Science \cite{PlotkinST91} literature.

Finding the decomposition required an understanding of network flow theory.
The ability to solve large problems hinges
on a fast network flow algorithm.
Classic augmenting paths algorithms
are far too slow for our purpose.
Goldberg's algorithm incorporates both
recent research within the worst-case model
and heuristics discovered by empirical studies
(in the spirit of \cite{DIMACS93}).

Also useful were Hardin, Sloane, and Smith's sphere covers
\cite{HardinSS94}.  These enabled us to start with better
uniform covers than we might have otherwise.
Finally, in prototyping and testing ideas, it helped
to have a pre-existing library of relevant high-level data types
and algorithms.  For this we used LEDA \cite{MehlhornN}.

Worst-case analysis did side-track us slightly.
Although worst-case analysis suggested that network flow would be the
bottleneck for large problems, it was not at all.
Ironically, as described in \S~\ref{sec:impl},
our first attempt to keep the flow problems small
by using equivalence classes backfired:
our original implementation used a single hashing data structure
to hold all the equivalence classes;
although the standard worst-case model suggests hashing is quite fast,
its incautious use slowed the solutions of large problems by a factor of 50
due to the lack of locality of reference.

In conclusion, our experience suggests that a successful approach
rested on theoretical understanding, but required that it be creatively adapted
to take advantage of the particular structure of our problem instances.

\section{Acknowledgements}
Thanks to Ken Steiglitz for introducing two of the coauthors
and to an anonymous referee for helpful suggestions.

\bibliographystyle{plain}
\bibliography{full,names,you,theory}
  
\end{document}